\def\year{2022}\relax
\documentclass[letterpaper]{article} % DO NOT CHANGE THIS
\usepackage{aaai22}  % DO NOT CHANGE THIS
\usepackage{times}  % DO NOT CHANGE THIS
\usepackage{helvet}  % DO NOT CHANGE THIS
\usepackage{courier}  % DO NOT CHANGE THIS
\usepackage[hyphens]{url}  % DO NOT CHANGE THIS
\usepackage{graphicx} % DO NOT CHANGE THIS
\usepackage{natbib}  % DO NOT CHANGE THIS AND DO NOT ADD ANY OPTIONS TO IT
\usepackage{caption} % DO NOT CHANGE THIS AND DO NOT ADD ANY OPTIONS TO IT
\DeclareCaptionStyle{ruled}{labelfont=normalfont,labelsep=colon,strut=off} % DO NOT CHANGE THIS
\usepackage{algorithm}
\usepackage{algorithmic}

\usepackage{amsmath}
\usepackage{amsfonts}
\usepackage{subcaption}
\usepackage{comment}

\usepackage{newfloat}
\usepackage{listings}
\floatstyle{ruled}
\newfloat{listing}{tb}{lst}{}
\floatname{listing}{Listing}
\title{A Review of Topological Data Analysis for Cybersecurity}
\author{
    Thomas Davies
}

\affiliations{
    The Alan Turing Institute\\
    The British Library, \\
    London, UK.\\
    tdavies@turing.ac.uk
}

\begin{document}

\maketitle

\begin{abstract}
In cybersecurity it is often the case that malicious or anomalous activity can only be detected by combining many weak indicators of compromise, any one of which may not raise suspicion when taken alone. The path that such indicators take can also be critical. This makes the problem of analysing cybersecurity data particularly well suited to Topological Data Analysis (TDA), a field that studies the high level structure of data using techniques from algebraic topology, both for exploratory analysis and as part of a machine learning workflow. By introducing TDA and reviewing the work done on its application to cybersecurity, we hope to highlight to researchers a promising new area with strong potential to improve cybersecurity data science.
\end{abstract}

\section{Introduction}

Topological Data Analysis (TDA) allows practitioners to study the topology of a dataset using mathematical tools that come with strong theoretical guarantees. Roughly, the topology of data is its global structure, meaning we can use TDA to analyse the ‘shape’ of datasets in ways that traditional data analysis methods cannot. In cybersecurity, many of the actions that threat actors take are individually not problematic; it is only when you consider the actions together that an issue becomes apparent \citep{vectorisation}. This makes cybersecurity data particularly well-suited to analysis by TDA, as it is capable of spotting anomalous patterns at a global level.

%In comparison to more traditional data analysis techniques which only rely on the geometry of the data, TDA allows analysts to view data through the lens of topology.
%Specifically, topological spaces are defined to be the same if they can be continuously deformed into one another. For example, the doughnut is topologically the same as a doughnut, which is the same as a mug (Figure 1). Note that each has one hole - the number of holes is one example of a topological invariant: a property of the space that doesn't change if two spaces are topologically identical. 
The primary techniques in TDA are Mapper and persistent homology. Mapper is an algorithm that can embed large high-dimensional datasets into a much smaller graph in a topology-preserving way, enabling its visualisation and analysis. Persistent homology produces a persistence diagram: a concise summary of the topology of a dataset which can be vectorised and used as input to other machine learning methods. There is an increasing body of work that shows TDA capable of performing at the state of the art on a large number of machine learning tasks. To give just one example, \citet{pmlr-v108-zhao20d} add persistent homology to graph neural networks to achieve the state of the art on many graph classification tasks.
%\begin{figure}
%\centering
%\end{figure}
Both strands of TDA have found applications in cybersecurity. Vast networks are common in this domain, and Mapper can reduce the size of large datasets in a principled way, allowing human operators to better visualise and understand the data they are working with. Indeed, the primary application of Mapper that we see in the literature is to reduce large datasets of network traffic into more interpretable graphs that can be viewed by human analysts. Clusters in the mapper graph are shown to correspond to types of anomaly \citep{darknet_top}, and proximity to potential anomalies within the mapper graph can alert analysts to traffic worth further investigation  \citep{enterprise}. Persistence diagrams have also been used with success on cybersecurity data. 
\citet{vectorisation} say that `little information can be gained from analysing individual records, [as] often the presence of intrusion behaviour or other anomalous activity can only be detected by looking at the aggregate behaviour of related records'. 
Persistence diagrams offer the ability to summarise the global behaviour of a large number of records, so their application to cybersecurity is very natural. This is demonstrated in the literature, with persistence diagrams being applied to successfully identify anomalies in network logs \citep{anom_ph} and, coupled with convolutional neural networks (CNNs), identify IoT devices from encrypted packet data \citep{passive}. More broadly, this relates to detecting change points. Indeed, there is an active research grant on TDA for Threat Detection\footnote{\url{https://www.nsf.gov/awardsearch/showAward?AWD_ID=1925346}} which has produced research on detecting change points topologically \citep{change_points}.

Before we can apply TDA we need to embed cybersecurity data in a way that is compatible with topological tools. This can either be by embedding into $\mathbb{R}^d$, or we can apply other tricks to embed directly into simplicial complexes -- the type of structure required for TDA. The choice of embedding has a large influence on the success of TDA, so we review different choices made in the literature. These range from general graph embedding techniques to those that are cybersecurity specific.

\begin{figure*}
\centering
\includegraphics[width=0.95\textwidth]{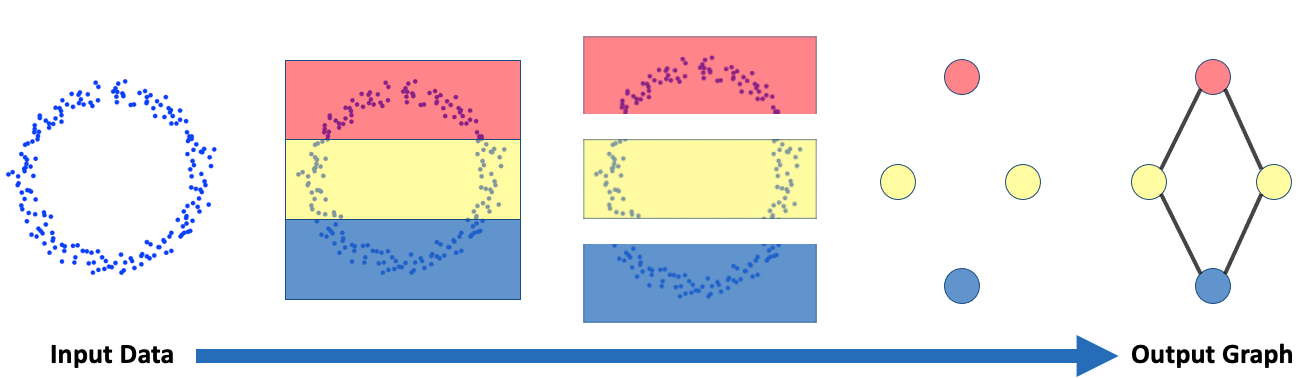}
\caption{In Mapper data is split up, clustered, then the resulting cluster centres are connected together based on shared nodes, leading to a graph that summarises the dataset in a topology-preserving way.}
\end{figure*}

The structure of this review is as follows. In Section 2 we give an introduction to Mapper and persistent homology, providing references for the interested reader. In Section 3 we summarise techniques to embed cybersecurity data as either vectors or graphs. In Sections 4 and 5 we review the literature on Mapper and persistent homology for cybersecurity respectively. In Section 6 we summarise research on graph metrics for cybersecurity that are closely tied to TDA. We conclude in Section 7, and in Appendix A we list datasets that may be of interest to readers.

\section{Overview of TDA}

Topological Data Analysis can be split into the Mapper algorithm and the persistent homology pipeline. Although not at first glance related, they both rely on the Nerve Theorem: a result in algebraic topology that guarantees a certain representation of a space preserves its topology \citep[Corollary 4G.3]{Hatcher}. This theoretical guarantee underpins TDA, allowing representations of data that preserve global structure. In the following we give a brief overview of TDA, but for more details see the textbook by \citet{comptopbook} or the lecture notes from \citet{vidit}.

\subsection{Mapper}

The Mapper algorithm \citep{mapper_paper} was the first algorithm to take advantage of a topological viewpoint of data. It can map complex high-dimensional datasets into lower-dimensional structures -- most commonly graphs -- in a topology-preserving way. This low-dimensional representation of the data is more interpretable than the data itself, and can be used to highlight areas of interest. Figure 1 gives an example of the whole Mapper process that we'd recommend referring to as we describe the algorithm. Given a dataset we split it up into overlapping areas. This is most commonly done with hypercubes, but can be done with any function (the domain of the function determines what structure you map onto). For the data contained in each hypercube  run a clustering algorithm; each cluster is a node in the output of the algorithm. When two clusters from different hypercubes share a point from the original data (which will happen, as the hypercubes overlap), add an edge between the nodes in the output of the algorithm. In this way you can map large high-dimensional datasets into a graph which the Nerve Theorem guarantees will preserve the structure of the initial dataset.

The results of Mapper depend heavily on the way you split your data and the clustering algorithm used, but experiments have shown Mapper to be very capable of providing new insight to data with sufficient parameter tuning, particularly when complemented with domain knowledge. To give some examples, \citet{mapp1} gave applications of Mapper to  tumours, voting, and player performance in the NBA. \citet{mapp2} used Mapper to identify a new subtype of breast cancer with high survival rates, a result that was one of the early indications of the efficacy of TDA.

%A spin-off company, Ayasdi, was founded with the Mapper algorithm at the core of its algorithm. Although less detail is offered on their workings, they use the Mapper algorithm for applications including protein analysis \citep{mapp3, mapp5}, traumatic brain injury biomarkers \citep{mapp4}, and infection tolerance \citep{mapp6, mapp7}.

%and a continuous map $f:X \to \mathbb{R}$. Let $I=\mathrm{range}(f|_X)$, and $\mathcal{S}$ be a partition of $I$. 
% Specifically, let $\mathcal{S}= \{I_j\}$ be a collection of intervals covering $I$, described by the intervals length $l$ and percentage overlap $p$. Then for each $I_j \in \mathcal{S}$, define $X_j = \{x \in X : f(x) \in I_j\}$, so that $\{X_j\}$ defines a cover of $X$. Cluster each $X_j$ (using a clustering method of your choice) to get clusters $\{X_{jk}\}$. Finally, you obtain a simplicial complex with vertices $\{X_{jk}\}_{j,k}$ and an edge from $X_{jk}$ to $X_{lm}$ whenever $X_{jk} \cap X_{lm} \neq \emptyset$.

\begin{figure*}
\centering
  \includegraphics[width=1\textwidth]{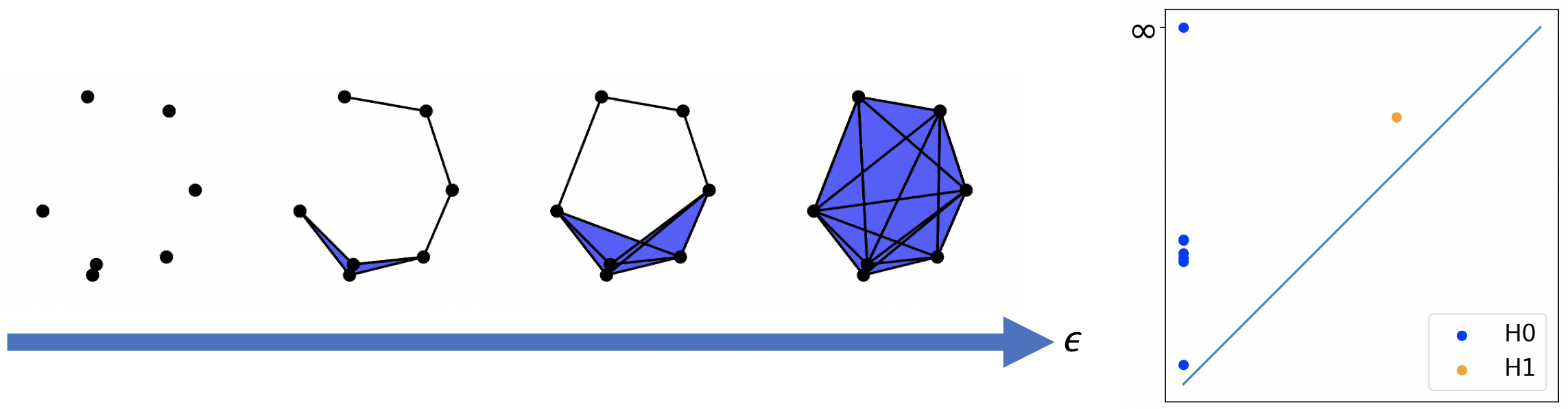}
\caption{On the left is a Vietoris-Rips filtration, shown with increasing values for $\epsilon$. On the right is the persistence diagram of the filtration. Each blue point is in the 0-persistence diagram, and represents a connected component. The orange point is in the 1-persistence diagram and represents the hole.}
\end{figure*}

\subsection{Persistent homology}

In the following we summarise how the persistent homology pipeline works. The key takeaway is that we end up with a (potentially vectorised) persistence diagram that concisely summarises the global structure of the inputted data. This can be used as-is or fed into a machine learning technique of your choosing for downstream tasks.

To compute the persistent homology we need a filtration of simplicial complexes. A simplicial complex is a collection of nodes, edges, triangles, and other higher dimensional equivalents, such that every simplex contains each of its constituent simplices (i.e., for any triangle in the complex, the complex also contains its edges and vertices), and the intersection of any two simplices in the complex is also a simplex in the complex. A filtration of simplicial complexes is a series of simplicial complexes $K_0, K_1, K_2, \dots$ such that $K_0 \subseteq K_1 \subseteq K_2 \subseteq \dots$ (see Figure 2). The parameter $\epsilon$ for complex $K_\epsilon$ is referred to as the time or scale of the filtration. We can build simplicial complexes from data in many ways. Given a point dataset, a common way is the $\epsilon$-Vietoris-Rips complex \citep{rips}. We add $k$ points as a simplex to the complex when they are pairwise within $\epsilon$ distance of each other. Letting $K_\epsilon$ be the $\epsilon$-Vietoris-Rips complex gives a filtration of simplicial complexes which we can use to compute the persistent homology. We can also create complexes in other ways. For example, in cybersecurity we may want to to build complexes from network data, connecting nodes that have had network connections with each other. In this case our scale parameter $\epsilon$ could be the time at which the network connection took place. This results in an equally valid filtration which we can use to compute the persistent homology.

For each simplicial complex $K_\epsilon$ we compute the $p$-homology group. This tells us the topology of $K_\epsilon$: the $0$-homology counts the number of connected components, the $1$-homology counts the number of holes, the $2$-homology counts the number of voids (e.g., the hole inside of a football), and so on \citep{toppers}. Thus we can use homology groups to describe the topology of $K_\epsilon$. However, we need to decide when computing the homology group which value of $\epsilon$ we want to consider. In the case of a Vietoris-Rips complex, small values of $\epsilon$ will result in disconnected points, whereas large values of $\epsilon$ will connect everything. The persistence trick circumvents that choice by allowing us to consider every $\epsilon$ simultaneously. We construct the $p$-persistent homology group by summing the $p$-homology groups over all $\epsilon$. This allows us to see how topological features persist throughout the filtration: simplices may enter the filtration at $K_i$ which cause a hole. More simplices may enter the filtration, say at $K_j$, that fill the hole. In this case, we refer to the birth of the hole as at time $i$, its death at time $j$, and we say it persists for $j-i$. In this way we can characterise the topological features present throughout the filtration. Some features may never die (for example, there will always be at least one connected component), in which case we say the death time is infinity.

Birth and death times for topological features offer a good way to visualise the persistent homology of a filtration. We can plot a diagram called the persistence diagram, in which the $x$-axis represents birth time, the $y$-axis represents death time, and each point is a topological feature within the filtration. An example is shown in Figure 2, in which the orange point is in the $1$-persistence diagram and represents the hole we can see within the filtration. The further above the diagonal a point is in the diagram, the longer it persists for, and the more likely it is to represent a topological feature rather than noise. There are distances defined on the space of persistence diagrams (e.g., the Wasserstein distance), but the space of persistence diagrams is not Hilbert. Therefore in order to conveniently integrate persistence diagrams into machine learning workflows we must embed them. Persistence diagrams can be vectorised by a large number of techniques, but probably most common is persistent images \citep{persimages}. This puts a Gaussian on each non-diagonal point of the persistence diagram and integrates the resultant surface over a grid to vectorise the diagram. Other common approaches are persistence landscapes \citep{bubenik2015statistical} or Betti curves \citep{rieck2017topological}, both of which end up with functional embeddings.

\section{Embedding cybersecurity data}

Cybersecurity data usually comes in forms that aren’t immediately amenable to TDA or, in fact, any machine learning algorithm. As such, choosing how to embed the data is an important step in any data science workflow for cybersecurity. Much of the data naturally has a graph structure, as many things that we consider live on a network. Some researchers choose to embed the data in a way that respects this graph structure, whereas others choose to disregard it; both options have had success in the literature. In this section we cover cyber-specific embeddings that both respect and disregard the graph structure, as well as more general graph embedding techniques.

In the TDA for cybersecurity literature, we most often see Mapper applied to vectors that are built in a graph-agnostic way, even when the initial data has a graph structure. This is unnecessary, as Mapper can be applied directly to graphs (Hajij, Rosen, and Wang 2018), but the tools to do so are less developed. In the case of persistent homology, again the majority of papers embed the data in a graph-agnostic way before computing the Vietoris-Rips complex. However, some approaches (Collins et al. 2020) use the existing network structure and create their filtration using features in the data, showing that such an approach is feasible.

\subsection{Graph-agnostic embeddings} \citet{vectorisation} proposed vectorising network log data by summing numeric fields and counting enumerated fields. They suggested the following feature vectors:
\begin{itemize}
    \item source IP, number of destination IP addresses;
    \item destination IP, number of failed access attempts;
    \item source IP, destination IP;
    \item destination perspective vector (consisting of destination IP, number of successful accesses, number of failed accesses, count of destination points, number of inbound bytes, number of outbound bytes).
\end{itemize}
This technique has been used in several pieces of anomaly detection research \citep{anomex1, enterprise}.

 \citet{anom_ph} proposed embedding NetFlow data into feature vectors that summarise either the cumulative number of packets sent or received by an IP of interest, the total number of packets sent or received by each IP, or the total number of packets sent or received by each IP to an IP of interest. These vectors are created over a window of a given length that advances over the whole dataset. The length of window and increment time are parameters that the authors vary. They show that the choice of window length and increment time can affect the performance on downstream tasks.
 %Over a sliding window of length $\delta$ starting at time $t$ they form a vector where the $i$th entry is either (i) the total number of packets sent/received by a given machine between time $t$ and $t+i$, (ii) the number of packets sent/received by machine $i$ between $t$ and $t+\delta$, or (iii) the total  number of
% In particular, let $\delta$ be the length of time we wish to consider, $s$ be the number of unique IPs on the network and $z_1,\dots,z_s$ be the unique IPs. For the first feature vector let $v \in \mathbb{R}^\delta$, with $(v)_i$ equal to the total number of number of packets sent or received by $y$ up to time $i$. The second technique gives a vector $v \in \mathbb{R}^s$ with $(v)_i$ equal to the number of packets sent/received by $z_i$. The third gives a vector $v \in \mathbb{R}^s$ with $(v)_i$ equal to the number of packets sent between $z_i$ and $y$.
 
\begin{figure}
\centering
\includegraphics[width=0.5\textwidth]{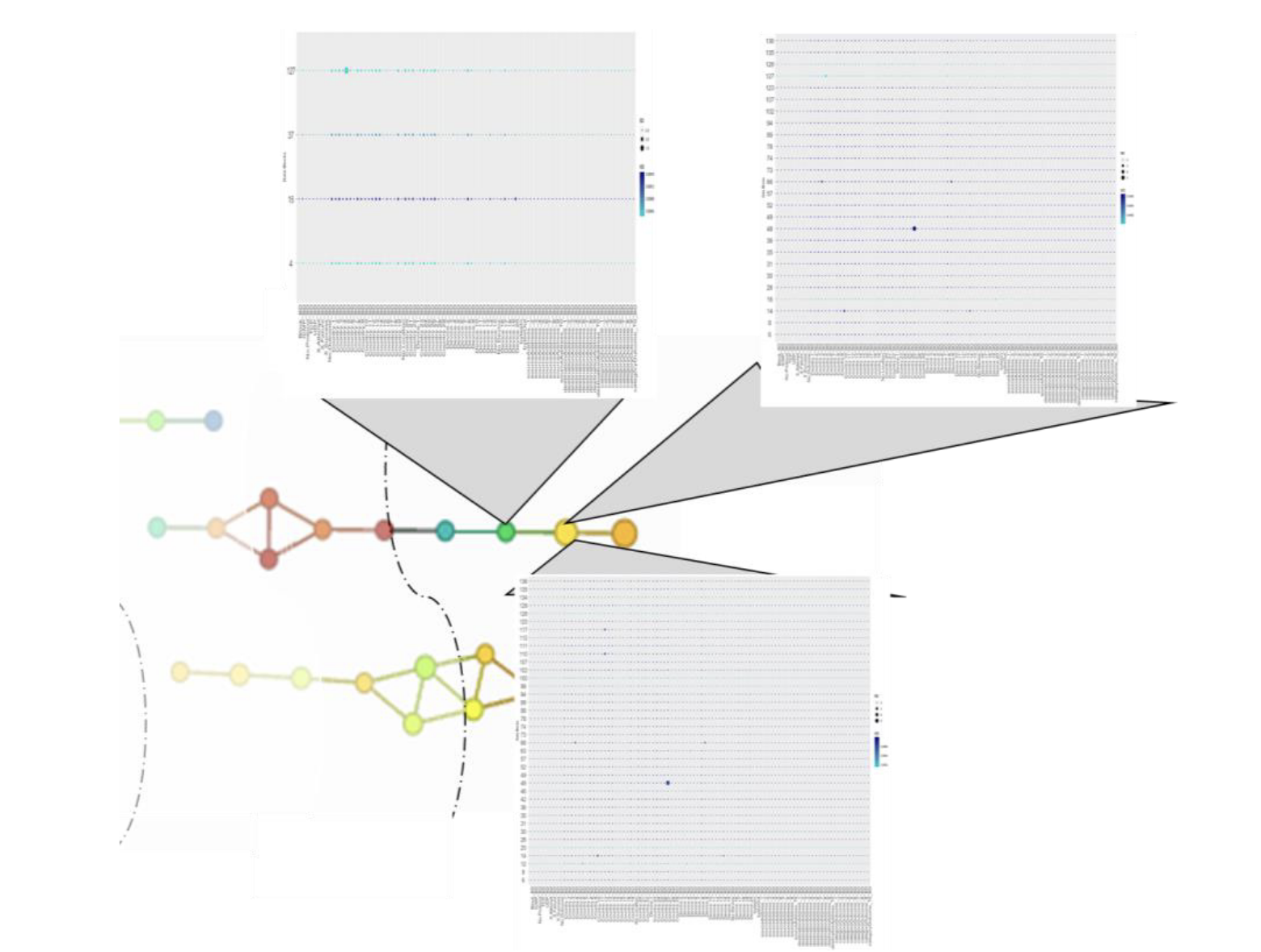}
\caption{Mapper applied to firewall logs. Nodes in close proximity to those flagged as potentially anomalous could be worth further investigation. Reproduced with permission from \citet{enterprise}.}
\end{figure}
 
\subsection{Event data as graphs} \citet{aksoy2019relative} considered a sliding window of 60 seconds that advanced 20 seconds at a time over event logs. For the events in each window they construct a graph -- the specific graphs they consider are listed below.
\begin{itemize}
    \item Authentication graphs are unweighted graphs built from authentication data that has source user/destination user as edges.
    \item Authentication failure graphs are as above, but restricted to failed authentication events. 
    \item Process graphs are built from start/stop records of processes. The graph consists of edges between computers and process names. 
    \item DNS graphs are built from DNS lookup events with edges from source computer to resolved computer.
    \item Flow graphs are built from the records of every network flow event. The edges are between the source computer and the destination computer. 
\end{itemize}

\citet{passive} used encrypted packet data to build a filtration. They picked a device of interest and connected it to the last $n$ devices that it exchanged packets with. They induced a filtration by the inter-packet arrival time, a metric that has previously been shown to contain discriminative information when the traffic is encrypted. Once all edges of a 2-simplex were present, they added the 2-simplex (Figure 6a). By doing so they avoided computing the expensive Vietoris-Rips complex, and instead used the natural structure of the data and a relevant feature to induce a filtration.

\subsection{Graph embeddings} Given the prevalence of graph-structured datasets in cybersecurity, we offer a brief review of techniques to embed graphs into $\mathbb{R}^d$.

\textbf{Skip-gram based models.} Skip-gram was introduced by \citet{skipgram0, skipgram} to improve word embeddings for NLP. Given a word from a sentence, the idea of skip-gram is to maximise the average log probability that your model can predict the next word. Node2Vec \citep{node2vec} and LINE \citep{tang2015line} adapted this for graph-structured data by maximising the probability that, given a node $v$, you can predict a node in the neighbourhood of $v$. Since the neighbourhoods over several hops can grow prohibitively large, they are sampled by random walk, with different adaptations using different sampling techniques.

\textbf{Graph neural networks.} GNNs learn to aggregate feature vectors across nodes by combining features with those of their neighbours.  Initially graph neural networks were classified as either spectral - aggregating neighbours in the spectral domain - or spatial - aggregating otherwise. However, true spectral filters require an expensive computation of the Fourier basis, so in general this was approximated, resulting in networks that were essentially spatial. As such, GNNs are now categorised as either convolutional, attentional, or message-passing. The difference lies in how information from the neighbourhood is aggregated. Convolutional networks aggregate with a linear sum using fixed edge weights as coefficients for a function of the node feature vectors \citep{kipf}. Attentional networks learn an attention function on edges that acts as the coefficient \citep{monti2016geometric}. Message-passing networks are the most general, learning one function on adjacent feature vectors that decides now to aggregate them  \citep{battaglia2016interaction}.

\begin{figure*}[ht!]
\centering
\includegraphics[width=1\textwidth]{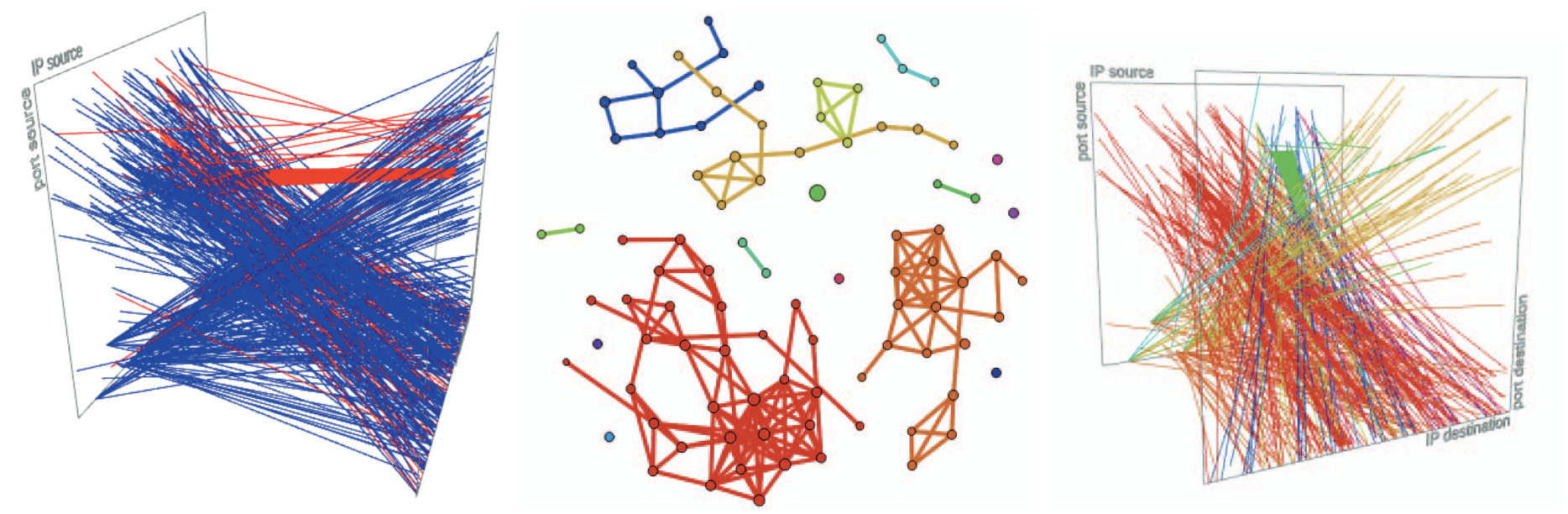}
\caption{Data collected by a network telescope is shown on the left. In the middle diagram is the Mapper graph of the data. Manual examination of the connected components showed that the large green dot is traffic attempting to exploit a known router vulnerability, the red component is traffic trying to access Telnet or SSH, the orange component is a sparse port scan from a single address, and the yellow component is a randomised scan and some noise. On the right the initial data is coloured according to the Mapper graph. We can see that the components are disorderly within the data, despite Mapper pulling out distinct attack types. Reproduced with permission from \citet{darknet_top}.}
\end{figure*}

\textbf{Embedding large graphs.} The techniques we have discussed so far cannot scale to large graphs, which is problematic as cybersecurity data can be very large. GraphSage \citep{graphsage} samples neighbours rather than fully computing the neighbourhood, and mini-batches, i.e., only updates the gradients for a limited number of nodes at a time. By doing so it enables large graphs (\textgreater100,000 nodes) to be processed with GNNs. ClusterGCN \citep{clustergcn} and GraphSAINT \citep{graphsaint} speed up GraphSage by removing some redundant computations. Facebook and Twitter have both also released methods for prcessing very large graphs. Twitter restricted large GNNs to a single layer \citep{sign} enabling huge computational speed-ups with much more parallelisation. However, the authors of this work suggest in a blog post\footnote{\url{https://blog.twitter.com/engineering/en_us/topics/insights/2021/simple-scalable-graph-neural-networks}} that it is best used as as a fast benchmark for other models, so the lack of depth does ultimately seem to effect the performance of the network. Facebook proposed Pytorch Big Graph \citep{pbg}. PBG works on multi-relational graphs: graphs where you can have different types of edges. It introduces several tricks to allow multi-relational GNNs to work on graphs with billions of nodes and trillions of edges.

\section{Mapper for cybersecurity}

In this section we review the literature on applications of Mapper to cybersecurity, splitting it by application domain. We find that it is most often used as an investigative and visualisation tool, although in some cases can provide quantitative information.

\textbf{Anomaly detection.} \citet{enterprise} applied Mapper to firewall logs from an enterprise-level organisation. The authors used a graph agnostic embedding technique developed by \citet{vectorisation} that we cover in Section 3 (counting categorical fields and summing numerical fields). After embedding the logs they computed the histogram matrix (HMAT) introduced by \citet{hmat}. This was intended to provide a way for analysts to visualise logs that may be anomalous over some block of time. \citet{enterprise} associated an HMAT with each node in the mapper graph, then anomalous nodes in the Mapper graph can flag to analysts that adjacent nodes may also be worth further investigation (Figure 3). However, this claim goes relatively untested, and the tool is framed more as an exploratory technique than something that can reliably find anomalous events. \citet{rizzo2020topological} similarly used Mapper to plot intrusion detection system data.

\textbf{Network telecopes.} Ranges of IPs with no active domains can be used to analyse traffic not directed towards real hosts. Since devices listening to this traffic are not real, any traffic to them is likely to be malicious \citep{darknet}. In the literature such traffic is called internet background radiation (IBR) and the listening devices are referred to as network telescopes. \citet{darknet_top} used Mapper on data from a network telescope to find port scans and DDoS attacks, outperforming traditional clustering algorithms when attacks don't cover the full range of possible ports or IPs. Of the many port scans identified by Mapper in the data, only a small proportion were found by the ruleset of the intrusion detection system Suricata, demonstrating that Mapper can outperform standard industry techniques in some cases. Mapper is also able to distinguish types of attack from complex data: examining the connected components of the Mapper graph of source/destination IP/port data from a network telescope showed that most components represent a distinct attack, including the attempted exploitation of a known router vulnerability (Figure 4).

\citet{Narita2021} also computed the Mapper graphs of network telescope data, but instead focused on describing the resultant graphs without attempting to link them to different attack types.

\textbf{Tor traffic detection.} If encrypted Tor traffic is detected as it passes through a network then connections can be immediately shut down and investigated. \citet{tor_enc} investigated Mapper as an unsupervised technique for visualising and classifying Tor traffic within a network. There is some success, with qualitative observations that Tor traffic tends to be on the extremes of the Mapper graph.

\begin{comment}
\begin{figure*}
     \centering
     \begin{subfigure}[b]{0.4\textwidth}
         \centering
         \includegraphics[width=\textwidth]{figs/portscan_fig.png}
         \caption{Portscans}
     \end{subfigure}
     \hfill
     \begin{subfigure}[b]{0.4\textwidth}
         \centering
         \includegraphics[width=\textwidth]{figs/ddos_fig.png}
         \caption{DDoS}
     \end{subfigure}
     \caption{Portscans and DDoS attacks captured by Mapper in network telescope data \citep{darknet_top}.}
\end{figure*}
\end{comment}

\textbf{Ransomware payments.} \citet{heist} computed a Mapper graph from Bitcoin transaction data. If a certain proportion of addresses that are within a node of the Mapper graph are known to receive ransomware payments then the remaining addresses have their risk scores increased. Repeating over the entire map and thresholding the acceptable risk gave a list of suspicious Bitcoin addresses. This approach outperformed DBSCAN and XGBoost at detecting suspicious addresses.

\textbf{Attack graphs.} An attack on a computer network consists of several different stages, which are categorised by the Mitre ATT\&CK\textregistered \hspace{1pt} framework as reconnaissance, delivery, exploitation, operation, data collection, and exfiltration. Such a sequence of actions naturally admits a graph structure, referred to as an attack graph. This graph is susceptible to analysis with topological techniques, which was the focus of work by \citet{huma}. They focused on context, attack patterns, and events, which combine to create an attack graph.
%They focus on APT (advanced persistent threat) attacks, which share six phases: reconnaissance, delivery, exploitation, operation, data collection, and exfiltration. 
They applied mapper to this attack graph in the hope that the reduced graph is able to be analysed by humans who can find repeated structures across attacks.

\begin{comment}
\begin{figure}
\centering
\includegraphics[width=0.4\textwidth]{figs/mapper.png}
\caption{Mapper graph of packet data within a network. Extremities of the graph tend to be Tor traffic \citep{tor_enc}.}
\end{figure}
\end{comment}

\section{Persistence for cybersecurity}

Persistence diagrams capture specific topological information about data which can be used independently for data analysis, or vectorised and fed into machine learning methods for further analysis and prediction. We see in this section that this can be used for anomaly detection and classification in far more quantitative ways than Mapper.

\textbf{Anomaly detection.} \citet{anom_ph} used the distance between persistence diagrams computed from NetFlow data to detect anomalies. Vectorising the NetFlow data with a sliding window over the data counting packets (as described in Section 3), the authors computed a baseline persistence diagram from a collection of feature vectors $B$. For every other vector $x$, they computed the Wasserstein distance between the persistence diagrams of $B$ and $B\cup \{x\}$. The larger the distance, the more topologically different the points are, and they found that spikes in topological dissimilarity are indicative of anomalies (Figure 5).

\textbf{Activity prediction.} \citet{gabdrak} summarised persistence diagrams with the Euler characteristic. Specifically, this is the alternating sum of the Betti numbers, which are approximated by the total persistence of the points in persistence diagrams, increasing by dimension. Given the levels of recent activity at a network switch (say, at an ISP), they wished to predict the future levels of activity. They have bits/minute data for 7 days. They split the data into two hour chunks and computed a persistence diagram for each two hour period. From that, they estimated the Euler characteristic of that period, as described above. They inputted the Euler characteristics into a neural network to predict future activity, but did not evaluate the success of their approach.

\begin{figure}
\centering
\includegraphics[width=0.5\textwidth]{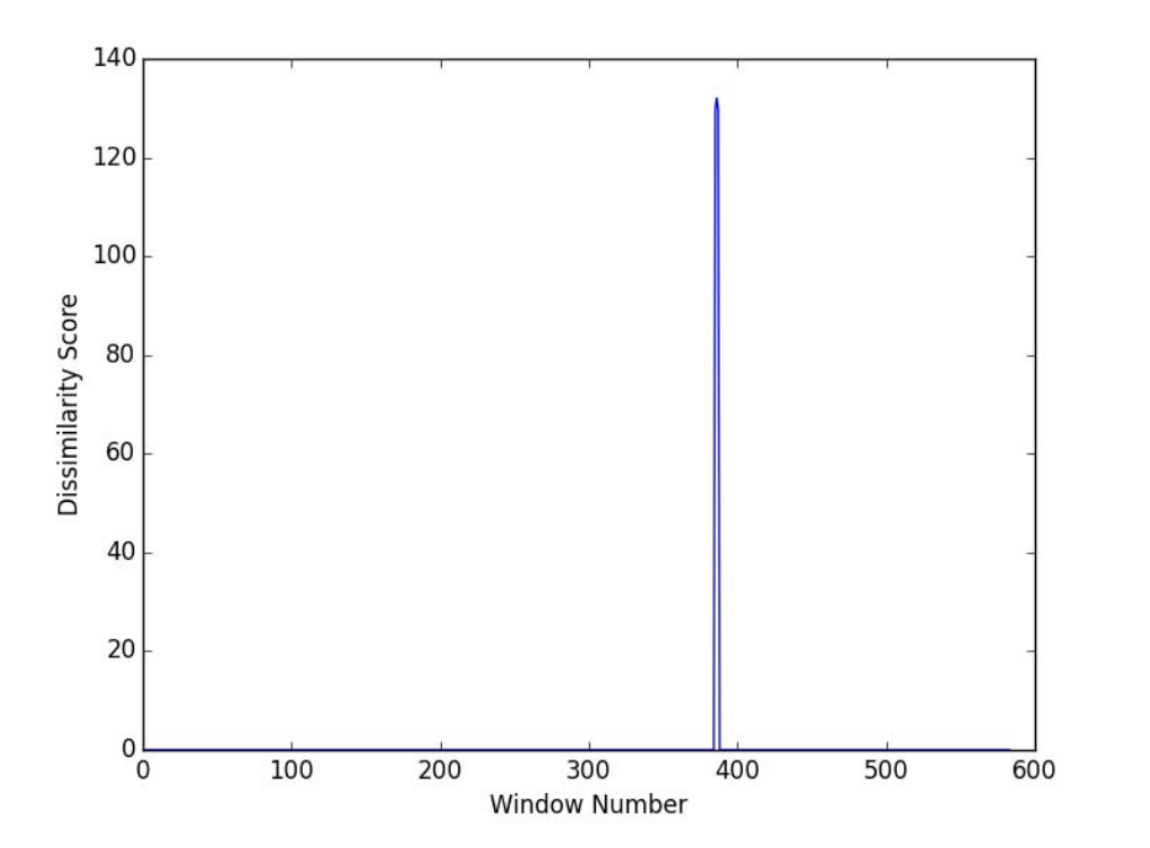}
\caption{The spike in topological dissimilarity indicates a predicted anomaly. In fact, this was a port scan. Reproduced with permission from \citet{anom_ph}.}
\end{figure}

%\textbf{Malware identification.} \citet{marchette} appear to also use the approximate Euler characteristic, in this case to identify malware. Details online are sparse, with the cited presentation slide being the only work available.

\textbf{IoT device fingerprinting.} \citet{postol2019time} showed that persistent homology works particularly well for classifying incomplete and noisy data from internet of things (IoT) devices on networks. They computed the persistence diagrams of embedded IoT network traffic, and used feature selection then logistic regression on functional embeddings of the diagrams to classify types of IoT devices (differentiating cameras, sensors, and multipurpose devices like tablets). With data over a longer period of time (months) this worked very well, but over shorter periods of time (days) it did not.

%It uses a summary of persistence diagrams computed from a sliding window over time series data from IoT traffic, and fed the resulting summaries into a classifier to evaluate how well persistent homology was capturing information about the class of IoT device . One interesting observation of the paper is that TDA did not do well at classifying devices over a short period of time (a few days), but when studying data over longer periods (on the scale of months) it performed much better. This implies that the global structure TDA is picking out can take time to appear in networked devices (although, we shall see a paper that succeeds on a smaller timescale next).

\citet{passive} also studied IoT data, only they used the natural network structure of the data, inducing a filtration with the inter-packet arrival time (IAT), a feature which has been previously shown to contain valuable information for traffic classification and anomaly detection \citep{uluagac2013passive}. By using the intrinsic structure of the data, they got rid of the need to compute the often prohibitively expensive Rips complex (Figure 6a). Next they computed the 1-persistent homology of the filtration. Finally they vectorised the persistence diagrams by mapping them to persistence images (Figure 6b), and used those to train a CNN. With these higher-dimensional topological features, they were able to identify IoT devices with high accuracy, recall, and precision.

%\citep{hagberg2014connected} study authentication graphs: bipartite graphs between users and computers, with authentication events forming edges. Although not topological, they study the graphs with graph theory, and the graphs they build could be studied topologically to look for anomalous behaviour.

\section{Graph metrics for cybersecurity}

\begin{figure*}
     \centering
     \begin{subfigure}[b]{0.46\textwidth}
         \centering
         \includegraphics[width=\textwidth]{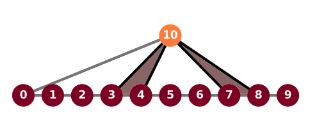}
         \caption{A simplicial complex built from packet data.}
     \end{subfigure}
     \hfill
     \begin{subfigure}[b]{0.46\textwidth}
         \centering
         \includegraphics[width=\textwidth]{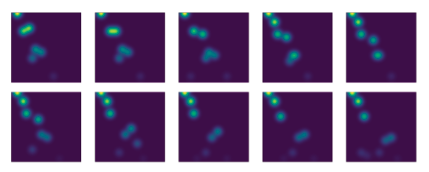}
         \caption{Examples of 1-persistence images of encrypted packet data.}
     \end{subfigure}
     \caption{The 1-persistent homology of packet data can be used to accurately fingerprint IoT devices on networks, even when the traffic is encrypted. Reproduced with permission from \citet{passive}.}
\end{figure*}

\citet{aksoy2020directional} introduced a new centrality measure for graphs based on the derivative of the graph Laplacian. The kernel of the Laplacian is exactly the 0-homology group, whilst the non-zero image is known to contain rich geometric descriptors of the graph. Therefore by using the Laplacian we capture both topological and geometric information about the network. Their centrality measure can detect synthetic anomalies that have been injected into the network data. Another measure to compare graphs is the relative Hausdorff distance. This offers a fast but nuanced way to compare two graphs based on their complementary cumulative degree histograms (CCDH). The CCDH of a graph $G$ is $(N(k))_{k=1}^\infty$, where $N(k)$ denotes the number of vertices in $G$ with degree at least $k$. Comparing the CCDHs gives us the relative Hausdorff distance, which \citet{aksoy2019relative} used to detect anomalies in graph sequences built from the Los Alamos dataset (see Appendix A). They built authentication graphs, authentication failure graphs, process graphs, DNS graphs, and flow graphs. Over longer time windows, they found that a spike in pairwise RH distance is indicative of a red-team event.

\section{Conclusion}

In this review we have introduced Mapper and persistent homology, and reviewed how they have been applied to cybersecurity problems. Mapper has mainly been used for visualisation, providing an effective way for analysts to get a better grasp of large datasets and suggesting data points for further investigation. However, it is hard to critically evaluate the success of these techniques, as the papers offer no robust feedback on how useful analysts find the visualisations. \citet{darknet_top} clustered the Mapper graph of network telescope data, showing that each cluster represents individual attack types. This demonstrates the potential of Mapper for unsupervised classification of anomalies in network data.

We have also seen that persistent homology can work on cybersecurity data. \citet{anom_ph} showed that topological dissimilarity can detect anomalous events taking place in network data using the 0-persistent homology: a spike in topological dissimilarity indicated port scans or DDoS attacks in the network logs. \citet{passive} built simplicial complexes directly from the intrinsic network structure in the data, training CNNs on the 1-persistence images. This demonstrated that higher order topological features can be used in cybersecurity, as they are able to very accurately fingerprint devices using just topological representations of encrypted network traffic.

%However, it did only consider the $0$-persistent homology, i.e., it only counted connected components. Their embedding technique means that an additional connected component will be introduced when there is a spike in the number of packets, such as when there is a DDoS or port scan on the network (as in the two examples given in the paper). Therefore, although this paper detects a topological change that is indicative of anomalous events, we should be sure to compare against simple baselines (such as counting packets) before using computationally intense topological tools. 

There is a lot of scope for further work in applying topological data analysis to cybersecurity. In particular, current literature on persistent homology demonstrates that it is able to use TDA's ability to concisely summarise global structure to perform valuable tasks in cybersecurity. We hope that this review brings more attention to TDA from people working on AI in cybersecurity.

\appendix
\section{List of datasets}

We have compiled a list of datasets that may be of interest to researchers in TDA for cybersecurity. Some of them are described in a review paper by \citet{datasets}.

\begin{itemize}
	\item \textbf{KDD Cup 1999} provides TCP connections, content features suggested by domain knowledge, and traffic features, along with labels for anomaly detection \citep{lippmann2000evaluating}. It is a very popular dataset for anomaly detection (although fairly outdated these days) and is available online at \url{http://kdd.ics.uci.edu/databases/kddcup99/kddcup99.html}. It is based on the data captured for the DARPA'98 IDS evaluation program. 
	
	\item \textbf{ECML-PKDD 2007} provides requests to web servers and was designed for identifying web attacks with machine learning \citep{dray2007web}. It is real-world data that has been manually processed and is available at \url{https://gitlab.fing.edu.uy/gsi/web-application-attacks-datasets}. 
	
	\item \textbf{CSIC 2010} is another dataset for web attacks and is based off real data from an e-commerce application. It is also available at \url{https://gitlab.fing.edu.uy/gsi/web-application-attacks-datasets}.

\item \textbf{ISOT Botnet and Ransomware} provides DNS queries for 9 botnets plus 19 malign applications, and 420GB of ransomware and benign program execution traces. It is available at \url{https://www.uvic.ca/ecs/ece/isot/datasets/botnet-ransomware/index.php}.

\item \textbf{CTU-13} consists of network logs for botnets, verified normal hosts, and unverified background noise \citep{garcia2014empirical}. It is available at \url{https://www.stratosphereips.org/datasets-ctu13}.

\item \textbf{ADFA} was compiled in 2013 to update intrusion detection datasets like KDD Cup that were over a decade old. The data consists of labelled system calls and is split into Linux, Windows, and Windows stealth attacks. It is available at \url{https://research.unsw.edu.au/projects/adfa-ids-datasets}.

\item \textbf{UNSW-NB15} is a comprehensive dataset that covers a variety of different attacks and is aimed at training and evaluating intrusion detection systems \citep{unsw}. It is available at \url{https://research.unsw.edu.au/projects/unsw-nb15-dataset}.

\item \textbf{DARPA Operationally Transparent Cyber} (OpTC) is a dataset of event logs from 1000 hosts that have a period of baseline activity, followed by having malware injected by a red team. It is available at \url{https://github.com/FiveDirections/OpTC-data}.

\item \textbf{LANL ARCS}, the Los Alamos National Lab Advanced Research in Cyber Systems dataset, consists of three separate datasets and is available at \url{https://csr.lanl.gov/data/}. The datasets are

\begin{itemize}
    \item Comprehensive, Multi-Source Cyber-Security Events, consisting of DNS and event data from across the LANL internal network, including labelled red team anomalies;
    \item Unified Host and Network Data Set, consisting of network data from the LANL internal network;
    \item User-Computer Authentication Associations in Time, which consists of around 700 million successful authentication events from the LANL enterprise network.
\end{itemize}

\item \textbf{BETH} is a recent anomaly detection dataset created by collecting over eight million DNS logs and kernel level events from 23 honeypots \citep{highnam763beth}. The data is real-life, contemporary and provided in a consistent format, and is available at \url{https://www.kaggle.com/katehighnam/beth-dataset}.
\end{itemize}

\section*{Acknowledgements}

This research was supported by the Defence and Security programme at the Alan Turing Institute, funded by the UK Government.

\bibliography{aaai22}

\end{document}